\documentclass[twocolumn,showpacs,preprintnumbers,amsmath,amssymb,nofootinbib]{revtex4-2}
\usepackage{comment} 
\usepackage{cancel}
\usepackage{graphicx}
\usepackage{dcolumn}
\usepackage{bm}
\usepackage[compact]{titlesec}
\usepackage{float}

\makeatletter
\DeclareRobustCommand{\cev}[1]{%
  {\mathpalette\do@cev{#1}}%
}
\newcommand{\do@cev}[2]{%
  \vbox{\offinterlineskip
    \sbox\z@{$\m@th#1 x$}%
    \ialign{##\cr
      \hidewidth\reflectbox{$\m@th#1\vec{}\mkern4mu$}\hidewidth\cr
      \noalign{\kern-\ht\z@}
      $\m@th#1#2$\cr
    }%
  }%
}
\makeatother
\usepackage{xcolor}

\definecolor{darkbrown}{RGB}{100, 20, 10}

\usepackage{hyperref}
\renewcommand{\arraystretch}{2.3}       
\newcommand{\half}{{{\textstyle\frac{1}{2}}}}
\newcommand{\quarter}{{{\textstyle\frac{1}{4}}}}
\newcommand{\be}{\begin{equation}}
\newcommand{\ee}{\end{equation} }
\newcommand{\beqa}{\begin{eqnarray} }
\newcommand{\eeqa}{\end{eqnarray} }
\newcommand{\ba}{\begin{array}}
\newcommand{\ea}{\end{array}}
\newcommand{\bpm}{\begin{pmatrix}}
\newcommand{\epm}{\end{pmatrix}}

\newcommand{\dis}{\displaystyle}

\newcommand{\rmd}{{\rm d}}
\newcommand{\rd}{{\rmd}}

\newcommand{\ODD}{\mathbf{O}(D,D)}

\newcommand\cF{{\cal F}}
\newcommand\cG{{\cal G}}
\newcommand\cH{{\cal H}}

\newcommand\cJ{{\cal J}}

\newcommand\cR{{\cal R}}

\newcommand{\Lp}{L_{\varphi}}

\newcommand{\trd}{{\bigtriangledown}}

\begin{document}


\title{Traversable wormhole for string, but not for particle}

\author{Hun Jang${}^{\,g,\scriptscriptstyle{B},\phi}$}
\author{Minkyoo Kim${}^{\,g}$}
\author{Hocheol Lee${}^{\,g}$}
\author{Jeong-Hyuck Park${}^{\,g}$}

\affiliation{${}^{g}$Department of Physics \& Center for Quantum Spacetime, Sogang University, Seoul 04107,  Korea}
\affiliation{${}^{\scriptscriptstyle{B}}$Yukawa Institute for Theoretical Physics (YITP), Kyoto University, Kyoto 606-8502, Japan}
\affiliation{${}^{\phi}$Research Institute of Basic Sciences (RIBS), Incheon National University,  Incheon 22012,  Korea}

\begin{abstract}
\noindent We propose a Lorentzian wormhole geometry characterized by a closed string massless sector with nontrivial $H$-flux and a scalar dilaton. In the string frame, the dilaton exhibits a negative kinetic term, enabling the existence of the wormhole. The geometry consists of three distinct regions. The middle region contains the throat, and its boundaries with the other two regions form non-Riemannian two-spheres, where a fundamental string becomes chiral, akin to a non-relativistic string. While point-particle geodesics are complete within each region and non-traversable across regions, strings perceive the geometry differently, allowing a chiral string to traverse freely.
\end{abstract}

                             
\maketitle


Chiral strings are fundamental constituents of string theory. In flat spacetime, an ordinary string arises as a superposition of chiral ({worldsheet-wise} left-moving) and anti-chiral (right-moving) modes. However, in highly curved spacetimes or in certain infinite limits of the spacetime metric, this pairing often breaks down, causing the string to become chiral~\cite{deVega:1987veo}. For example, chiral strings appear at black hole horizons, illuminating the microscopic origin of black hole entropy~\cite{Kogan:1986yd}; at cosmological orbifold singularities, where they prevent divergences in scattering amplitudes~\cite{Berkooz:2002je}; 
in worldsheet scattering theory, where they correspond to the fundamental asymptotic states~\cite{Hofman:2006xt} or {ambitwistor strings~\cite{Mason:2013sva}};
and in the non-relativistic limit of flat spacetime~\cite{Gomis:2000bd,Danielsson:2000gi,Gomis:2005pg}, which generalizes  to the recently explored  Newton--Cartan strings~\cite{Christensen:2013lma,Hartong:2015zia,Harmark:2017rpg,Harmark:2018cdl,Bergshoeff:2018yvt,Bergshoeff:2019pij,Harmark:2019upf,Bergshoeff:2021bmc,Oling:2022fft,Hartong:2022lsy}. Furthermore, in the geometric framework of double field theory, the entanglement of left- and right-moving modes can condense to produce a Riemannian spacetime from non-Riemannian pregeometry~\cite{Park:2020ixf}, identifying the metric as a Nambu–Goldstone boson~\cite{Berman:2019izh}, in line with earlier insights~\cite{Misner:1973prb}. Chiral closed strings remain localized in spacetime, whereas chiral open strings attach to null branes~\cite{Kogan:2001nn}. In this Letter, we introduce a wormhole as another example of a background that admits freely traversing chiral strings.

A Lorentzian wormhole is among the earliest solutions in General Relativity (GR), connecting distinct flat regions of spacetime~\cite{Flamm,Einstein:1935tc}. Its realization, stability, observability, and traversability---as well as its relationship to quantum entanglement---have been longstanding and active areas of research~\cite{Misner:1957mt,Morris:1988cz,Morris:1988tu,Giddings:1989bq,Visser:1995cc,Kim:1997jf,James:2015ima,Lobo:2017cay,Dai:2019mse,DeFalco:2020afv,Maldacena:2013xja,Gao:2016bin,Maldacena:2018gjk,Maldacena:2020sxe}. In particular, traversability depends on specific wormhole criteria~\cite{Morris:1988cz}, including the flare-out conditions~\cite{Lobo:2017cay} which necessitate violations of a {null energy  condition}.

As a leading candidate for  quantum  gravity, string theory naturally raises the question: \textit{Can traversable Lorentzian  wormholes exist within string theory without invoking exotic matter?}  It is the purpose of the present Letter to  propose a Lorentzian wormhole  within the context of string theory at leading order in $\alpha^{\prime}$  and to show its traversability  by chiral strings but not by particles nor ordinary strings. Our wormhole solution corresponds to  a   pure  Neveu--Schwarz Neveu--Schwarz  (NS-NS)  geometry   that does not require any (exotic) extra matter.  The gravitational action we assume  is   the renowned    low-energy effective action of the  NS-NS  string massless sector comprising a  metric, $B$-field, and scalar dilaton, \textit{i.e.~}$\{g_{\mu\nu},B_{\mu\nu},\phi\}$:
\be
\displaystyle{\int}\rmd^{D}x\,\sqrt{-g}e^{-2\phi}\!
\big(R+4\partial_{\mu}\phi\partial^{\mu}\phi-\textstyle{\frac{1}{12}}H_{\lambda\mu\nu}H^{\lambda\mu\nu}\big)\,,
\label{Reff}
\ee
of which the Euler--Lagrange equations lead to
\be
\ba{rll}
R_{\mu\nu}+2\trd_{\mu}(\partial_{\nu}\phi)-\quarter H_{\mu\rho\sigma}H_{\nu}{}^{\rho\sigma}
&=&0\,,\\
\half e^{2\phi}\trd^{\rho}\!\left(e^{-2\phi}H_{\rho\mu\nu}\right)&=&0\,,\\
R+4\trd_{\mu}(\partial^{\mu}\phi)-4\partial_{\mu}\phi\partial^{\mu}\phi-\textstyle{\frac{1}{12}}H_{\lambda\mu\nu}H^{\lambda\mu\nu}&=&0\,.
\ea
\label{EoM3}
\ee
Here $H_{\lambda\mu\nu}$ is the field strength of the $B$-field, or $H$-flux.  Although superstring theory is formulated in ten dimensions, we focus on a four-dimensional external spacetime (with $D=4$), implicitly leaving the detailed treatment of compactified  (Ricci flat) internal dimensions aside.

{Double Field Theory (DFT), initiated in \cite{Siegel:1993xq,Siegel:1993th} and \cite{Hull:2009mi,Hull:2009zb,Hohm:2010jy,Hohm:2010pp}, provides a framework in which  the entire Lagrangian~(\ref{Reff})  transforms into an $\ODD$-symmetric   generalized scalar curvature~\cite{Jeon:2010rw,Jeon:2011cn}.   Additionally,  the three equations of motion~(\ref{EoM3}) are unified into a single expression with $\ODD$ vector indices, 
${G_{AB}=0}$, representing  the vanishing of the   DFT Einstein curvature~\cite{Park:2015bza}, or equivalently,  the energy-momentum tensor via   the DFT Einstein equation~\cite{Angus:2018mep}:
\be
G_{AB}=T_{AB}\,.
\label{EDFE}
\ee
In this context,  our wormhole configuration, being a pure NS-NS geometry, corresponds to a vacuum solution of the  unified equation~(\ref{EDFE}). While our analysis is primarily conducted within the  Riemannian  framework, such as (\ref{EoM3}), without fully engaging with the DFT formalism underlying (\ref{EDFE}) (for a recent review, see \cite{Park:2025ugx}), certain conceptual insights from the DFT perspective---particularly regarding singularities,  energy conditions, and the stringy traversability---are unavoidable. }

\section*{NS--NS Wine-Glass Wormhole} 
The wormhole geometry we propose is a two-parameter family of solutions to (\ref{EoM3}) and is  traceable to the work~\cite{Burgess:1994kq} by Burgess, Myers, and Quevedo who obtained more general three-parameter family of solutions  by   performing  $\mathbf{SL}(2,\mathbb{R})$ S-duality rotations of a dilaton–metric solution in Einstein frame.  The three-parameter  solutions were later re-derived~\cite{Ko:2016dxa} as the  most general  spherically symmetric vacuum solutions to the DFT  Einstein equation~(\ref{EDFE}), by analogy with Schwarzschild geometry of GR. 
Without further ado, we spell the  (horizonless)  solution:
\be
\ba{c}
\rd s^{2}=\dis{\frac{-\rd t^{2} +\rd y^{2}}{\cF(y)}}+\cR(y)^{2}\left(\rd\vartheta^{2}+\sin^{2\!}\vartheta\,\rd\varphi^{2}\right)\,,\\
H_{(3)}=h\sin\vartheta\,\rd t\wedge\rd\vartheta\wedge\rd\varphi \,,\qquad
e^{2\phi(y)}=\dis{\frac{1}{\left|\cF(y)\right|}}\,,
\ea 
\label{SOL}
\ee
where $\cF$  and $\cR$ (areal radius) are functions of  $y$,  
\be
\ba{ll}
\cF(y)=\dis{\frac{(y-b_{-})(y-b_{+})}{y^2+\frac{1}{4}h^2}}\,,\quad&\quad
\cR(y)=\sqrt{y^{2}+\frac{1}{4}h^{2}}\,.
\ea
\label{FR}
\ee
The geometry has two   real free parameters, nontrivial  ${b\neq0}$ and electric $H$-flux $h$, in terms of which we   set 
\be
\ba{lll}
\gamma_{\pm}=\frac{1\pm\sqrt{1-h^{2}/b^{2}}}{2}\,,\quad&\quad b_{+}=-b\gamma_{+}\,,\quad&\quad b_{-}=b\gamma_{-}\,,
\ea
\ee
to acquire $b_{+}+b_{-}=-b\sqrt{1-h^2/b^2}$ and $b_{+}b_{-}=-\frac{1}{4}h^2$. 
For the solution to be real, we  require $h^{2}\leq b^2$. While the vanishing  limit $b\rightarrow 0$  may reduce the geometry  to a flat Minkowskian spacetime, when $b$ is strictly positive $b>0$ we have $b_{+}<0\leq b_{-}$ and if $b<0$ we get the opposite ordering, $ b_{-}\leq 0<b_{+}$. Obviously  we note 
\be
\cR(y)=\cR(-y)\,\geq\,\half\left|h\right|\,.
\label{Rpm}
\ee
The function $\cF(y)$  vanishes at the points of $y=b_{+}$ and $y=b_{-}$. Between them  it is   negative   and outside positive.  Thus, in the intermediate interval, $t$ and  $y$ become spatial and temporal  coordinates respectively.  Though  the areal radius is parity symmetric~(\ref{Rpm}),   the metric component $\cF(y)$ is  generically not,  except the case of  saturation, ${h^2=b^2}$. Therefore, in general one cannot identify $y$ with $-y$ to perform a $Z_{2}$-orbifolding, which would be necessary for the full realisation of the flat spacetime after taking the limit, $\left|b\right|=\left|h\right|\rightarrow0$.   We are led to set the range of the $y$-coordinate to be all real numbers, $y\in{\mathbb{R}}$. The geometry then consists of two separate,  asymptotically flat spacetime letting $\cF(y)\rightarrow 1$, one by $y\rightarrow\infty$ and  the other by $y\rightarrow-\infty$, which are to be  connected by a wormhole.  The minimum of the areal radius~(\ref{Rpm}) is assumed at $y=0$ which we identify as  the throat of the wormhole.  In fact,  from
\be
\ba{ll}
\scalebox{1.3}{$\frac{\rd \cR}{\rd y}=\frac{y}{\sqrt{y^2+\frac{1}{4}h^2}}$}\,,\qquad&\quad
\scalebox{1.3}{$\frac{\rd^{2}\cR}{\rd y^{2}}=\frac{\frac{1}{4}h^{2}}{(y^2+\frac{1}{4}h^2)^{3/2}}>0$}\,,
\ea
\ee
as long as the electric $H$-flux is nontrivial, $h\neq0$,  a  flare-out condition is  satisfied   in terms of the $y$-coordinate,
\be
\ba{ll}  
\scalebox{1.2}{$\left.\frac{\rd \cR}{\rd y}\right|_{y=0}=0\,,$}\qquad&\quad
\scalebox{1.2}{$\left.\frac{\rd^{2}\cR}{\rd y^{2}}\right|_{y=0}=\frac{2}{\left|h\right|}\,>\,0\,.$}
\ea
\label{FlareOut1}
\ee

\textit{---Embedding  Diagram  into an Ambient Spacetime.}
Following a well-known prescription~\cite{Lobo:2017cay},  we  embed the wormhole geometry into an ambient spacetime,
\be
\rd \hat{s}^{2}=\frac{-\rd t^{2}}{\cF}\pm \rd z^2+\rd\cR^2+\cR^{2}\!\left(\rd\vartheta^{2}+\sin^{2\!}\vartheta\,\rd\varphi^{2}\right)\,,
\label{ambient}
\ee
through a pair of functions in $y$:  $\cR(y)=\sqrt{y^2+\frac{1}{4}h^2}$ and  $z(y)$ satisfying 
\be
\!\frac{\rd z}{\rd y}\scalebox{1}{$=\!\sqrt{\pm\!\left[\frac{1}{\cF}-\left(\!\frac{\rd\cR}{\rd y}\right)^2\right]}\!=\!\sqrt{\pm\!\left[\frac{-b\sqrt{1-h^2/b^2}y^3+\frac{3}{4}h^2y^2+\frac{1}{16}h^4}{(y-b_{+})(y-b_{-})(y^2+\frac{1}{4}h^2)}\right]}$}\,.
\label{dzdy}
\ee
The sign in (\ref{ambient}) and (\ref{dzdy})  must be chosen to ensure the realness of  the square roots and hence the embedding is inevitably piecewise. We further set the integral constant to fix $z=0$ at $y=0$.  

\begin{figure}[H]
\centering
\vspace{-8pt}
\includegraphics[width=0.48\linewidth]{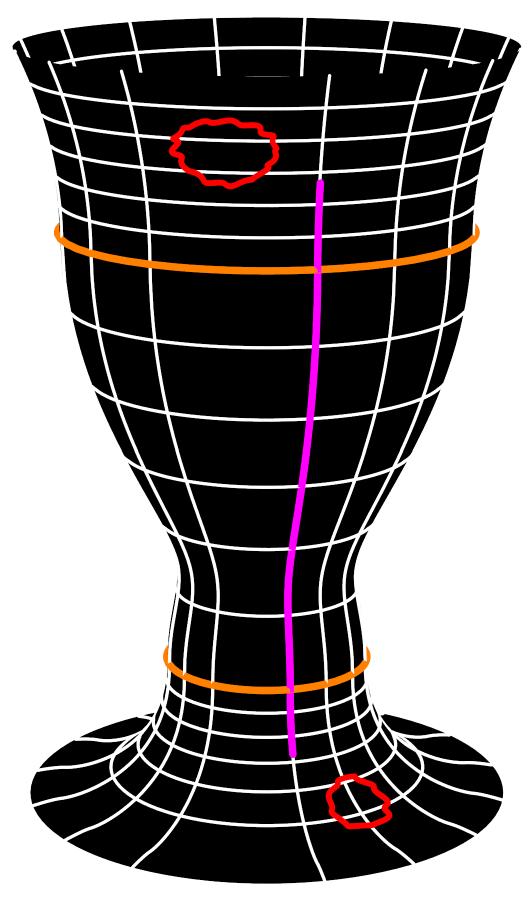}\vspace{-5pt}
\caption{{{Asymmetric, `wine-glass' shaped wormhole in an ambient space, with the choice  of $\,{b<h<0}$.   Non-traversing ordinary strings from (\ref{cGyff}) and traversing chiral strings from (\ref{transSOL})---either trajectory of (\ref{ytchiral}) or snapshot of (\ref{ellipsoid})---are  colored in red and pink  respectively.  The  Riemann-wise singular but DFT-wise regular non-Riemannian points   at ${y=b_{\pm}}$ are colored in orange.  }} }
	\label{FIGEmbedding}
\end{figure}

\textit{i)} Generically for  $b^2>h^2>0$,  the sign should change at three points,  $y=b_{-}$, $y=b_{+}$, and
\be
y=\,-\,\frac{\,h^2+\sqrt[3]{4h^{4}b_{+}^{2}}+\sqrt[3]{4h^{4}b_{-}^{2}}\,}{4(b_{+}+b_{-})}\,.
\ee
Asymptotically for large $\cR$ as $y\rightarrow\pm\infty$, we note
\be
z\,\sim\,\pm 2(b^2-h^2)^{1/4}\sqrt{\cR}\,.
\ee
This  supplements  the throat region depicted in FIG.\,\ref{FIGEmbedding}.\vspace{3pt}
\textit{ii)} When ${b^2=h^2>0}$, the sign  changes twice: at $y=b_{+}$ and at $y=b_{-}$    in accordance  with $\cF$. The signature of the ambient spacetime~(\ref{ambient}) remains everywhere Minkowskian. Asymptotically we get 
\be
z\,\sim\,\pm \frac{\sqrt{3}}{2}\left| b\right|\ln{\cR}\,.
\ee
The flare-out condition is normally   addressed in terms of the ambient spacetime coordinate $z$~\cite{Lobo:2017cay}.  By construction from (\ref{dzdy}), $z(y)$ is a monotonically increasing function of $y$, though its derivative diverges at $y=b_{\pm}$. Yet, at the throat of $z=y=0$, from (\ref{dzdy}) with the lower minus sign  chosen, it features regular behaviour:  ${\frac{\rd z(0)}{\rd y}=1}$ and  $
\frac{\rd^{2}z(0)}{\rd y^{2}}=2(b/h^{2})\sqrt{1-h^2/b^2}$. 
Thus, from (\ref{FlareOut1}) and through simple chain rule, we reassure  the flare-out condition  in terms of  the   ambient $z$-coordinate too, \textit{c.f.~}(\ref{FlareOut1}),
\be
\ba{l}  
\left.\frac{\rd \cR}{\rd z}\right|_{z=0}=\left.\big(\frac{\rd z}{\rd y}\big)^{-1}
\frac{\rd \cR}{\rd y}\,\right|_{y=0}=0\,,\\
\left.\frac{\rd^{2}\cR}{\rd z^{2}}\right|_{z=0}\!=
\left.\big(\frac{\rd z}{\rd y}\big)^{-2}\!\left[
\frac{\rd^{2} \cR}{\rd y^{2}}-\big(\frac{\rd z}{\rd y}\big)^{-1}\frac{\rd^{2}z}{\rd y^{2}}
\frac{\rd \cR}{\rd y}\right]\,\right|_{y=0}
\!=\frac{2}{\left|h\right|}>0\,.
\ea
\label{FlareOut2}
\ee

\begin{figure}[H]
\centering
\includegraphics[width=0.82\linewidth]{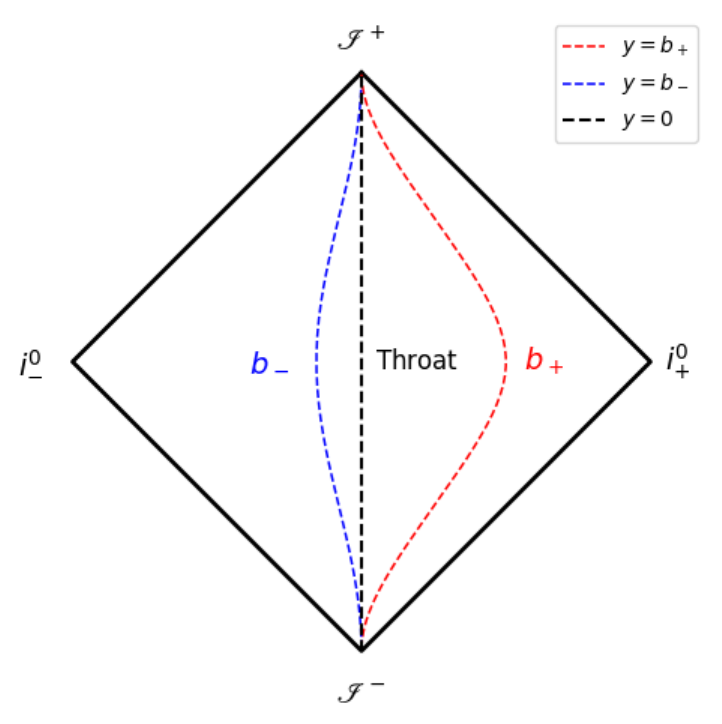}
\caption{{{Penrose Diagram  of the Wormhole Geometry for the choice of $b=-5/4$ and $h=1$ hence $b_{+}=1$ and $b_{-}=-1/4$.  While the temporal infinities $t\rightarrow\pm\infty$ are as usual denoted by $\cJ^{\pm}$,  the spatial infinities of $y\rightarrow+\infty$ and $y\rightarrow -\infty$  are denoted by ${i}^{0}_{+}$ and ${i}^{0}_{-}$ respectively.  In particular, the chiral string of (\ref{ytchiral}) traverses the wormhole at a $45$-degree angle. }}} 
	\label{FIGPenrose}
\end{figure}

\textit{---Penrose Diagram of the Wormhole.} In terms of compactified light-cone coordinates,
\be
\ba{ll}
U = \arctan (t-y)\,,\quad&\qquad V = \arctan (t+y)\,,
\ea
\ee
The wormhole metric~(5)  gives
\begin{equation}
 \dis{\frac{-\rd t^{2} +\rd y^{2}}{\cF(y)} = -\frac{\rd U\rd V}{\,\cos^2 U\,\cos^2 V\,\mathcal{F}\left(\frac{\tan V-\tan U}{2}\right)}\,.}
\end{equation}
We depict the corresponding Penrose diagram in FIG.\,\ref{FIGPenrose}.

{\textit{---GR Singularity   Identified As DFT Regularity.}  Within the conventional framework of Riemannian differential geometry, the wormhole exhibits curvature singularities at the points  $y=b_{\pm}$: 
\be
R=\scalebox{1.5}{$-\,\frac{\,2b^2(y^2+\frac{1}{4}h^2)^2+3h^2(y-b_{+})^2(y-b_{-})^2}{2(y-b_{+})(y-b_{-})(y^2+\frac{1}{4}h^2)^3}$}\,.
\ee
In contrast, from the perspective of DFT, the wormhole geometry is everywhere regular. The fundamental fields of DFT are the generalized metric $\cH_{AB}$ and the $\ODD$-singlet dilaton $d$.   Their parametrization for an NS-NS geometry is given by
\be
\ba{ll}
\cH_{AB}=\scalebox{0.9}{$\left(\ba{cc}g^{-1}&-g^{-1}B\\Bg^{-1}&{~g-Bg^{-1}B}\ea\right)$}\,,\quad&~\quad e^{-2d}=\sqrt{-g}e^{-2\phi}\,.
\ea
\label{PARA}
\ee
For the present wormhole geometry~(\ref{SOL}),  the inverse metric $g^{-1}$ and the $\ODD$-singlet dilaton $d$ remain non-singular everywhere, with $e^{-2d}=\cR(y)^2\sin\vartheta$. Furthermore,   by choosing the  $B$-field to include a pure gauge term:
\be
B_{(2)}=h\cos\vartheta\,\rd t\wedge\rd\varphi \,+\, \frac{\rd t\wedge\rd y}{\cF(y)}\,,~~\quad\rd B_{(2)}=H_{(3)}\,,
\label{Bfield}
\ee
it is  ensured that all components of  $\cH_{AB}$ in (\ref{PARA}) remain  finite and  regular,  as the  negative  powers of $\cF(y)$ are fully canceled~\cite{Morand:2021xeq}.  Only the  positive powers appear as seen from
\be
\ba{l}
g^{-1}=
\left(\ba{cccc}
~-\cF~&~0~&~0~&~0~\\
0&\cF&0&0\\
0&0&\frac{1}{\cR^{2}}&0\\
0&0&0&\frac{1}{\cR^{2}\sin^2\vartheta}
\ea\right)\,,
\\
Bg^{-1}=\big(-g^{-1}B\big)^{T}=\left(\ba{clcc}
~0~&1~~~~&0~&\frac{h\cos\vartheta}{\cR^{2}\sin^2\vartheta}\\
1&0&0&0\\
0&0&0&0\\
h\cos\vartheta\cF&0&0&0
\ea\right)\,,
\ea
\label{gB}
\ee
and
\be
\ba{l}
g-Bg^{-1}B\\
=\left(\ba{cccc}
\frac{h^2\cos^2\vartheta}{\cR^{2}\sin^2\vartheta}&0&0&0\\
0&0&0&-h\cos\vartheta\\
0&0&\cR^2&0\\
0&-h\cos\vartheta&0&\quad\scalebox{0.9}{$\cR^2\sin^2\vartheta-h^2\cF\cos^2\vartheta$}
\ea\right)\,.
\ea
\label{gBgBd}
\ee

In DFT, all geometric quantities, including curvatures, are defined solely by the fundamental fields  $\{\cH_{AB},d\}$ rather than $\{g_{\mu\nu},B_{\mu\nu},\phi\}$~\cite{Jeon:2010rw,Jeon:2011cn,Park:2015bza}. Consequently, the wormhole does not exhibit any DFT curvature singularity.  As the $B$-field gauge transformation is part of doubled diffeomorphisms, the curvature singularity within Riemannian geometry is identified as a coordinate singularity in DFT~\cite{Morand:2021xeq}. Indeed, the wormhole solution renders the $\ODD$-symmetric  Einstein curvature trivial in (\ref{EDFE}), and thus, analogous to GR,  both the  DFT  scalar  and  Ricci curvatures  vanish trivially, without any singularity.

At  ${y=b_{\pm}}$,  $\cF(y)$ vanishes,  causing  the upper-left block of the generalized metric (corresponding to $g^{-1}$ in (\ref{PARA})) to    become degenerate, preventing the definition of an invertible Riemannian metric. The DFT geometry transitions to a non-Riemannian regime, similar to that of  the non-relativistic chiral string theories~\cite{Gomis:2000bd,Christensen:2013lma,Hartong:2015zia,Harmark:2017rpg,Harmark:2018cdl,Bergshoeff:2018yvt,Bergshoeff:2019pij,Harmark:2019upf,Bergshoeff:2021bmc,Oling:2022fft,Hartong:2022lsy}.  DFT can describe such non-Riemannian geometries as consistent chiral string backgrounds~\cite{Lee:2013hma,Ko:2015rha,Park:2016sbw,Morand:2017fnv,Blair:2019qwi,Cho:2019ofr,Park:2020ixf}.  From the DFT perspective, the wormhole geometry is everywhere finite and regular. \\
\vspace{3pt}
}

{\textit{---Null Energy Condition (NEC).}  In GR, the null energy condition is equivalent to the null  convergence condition (NCC)}  which  requires  $R_{\mu\nu}k^{\mu}k^{\nu}\,\geq\,0$ for all future-pointing null vector fields~$k^{\mu}$~\cite{Lobo:2017cay,Parikh:2014mja}.  
For the NS-NS gravity~(\ref{Reff}), from (\ref{EoM3}),  the Ricci curvature decomposes on-shell   into dilaton and $H$-flux contributions:
\be
R_{\mu\nu}=-2\trd_{\mu}(\partial_{\nu}\phi)+\quarter H_{\mu\rho\sigma}H_{\nu}{}^{\rho\sigma}\,.
\ee
In the wormhole geometry~(\ref{SOL}), each term becomes diagonal when treated as  $4\times 4$ matrices. Contracting with a  null radial vector  $k^{\mu}=(1,1,0,0)$, we obtain
\be
\ba{rll}
R_{\mu\nu}\,k^{\mu}k^{\nu}\!&\!=\!&\!\scalebox{1.05}{$-\,\frac{4(b_{+}+b_{-})y(y^{2}-\frac{1}{2}h^2)+5h^2(y^2-\frac{1}{20}h^2)}{2\cR(y)^4(y-b_{+})(y-b_{-})}\,,$}\\
-2\trd_{\mu}(\partial_{\nu}\phi)\,k^{\mu}k^{\nu}\!&\!=\!&\!\scalebox{1.05}{$
-\,\frac{2(b_{+}+b_{-})y(y^{2}-\frac{3}{4}h^2)+3h^2(y^{2}-\frac{1}{12}h^2)}{\cR(y)^4(y-b_{+})(y-b_{-})}
\,,$}\\
\quarter H_{\mu\rho\sigma}H_{\nu}{}^{\rho\sigma}\,k^{\mu}k^{\nu}\!&\!=\!&\!
\frac{h^{2}}{2\cR(y)^{4}}\,>\,0\,.
\ea
\ee
{This demonstrates that the $H$-flux always provides a positive contribution to the NCC, whereas   the dilaton, with its negative kinetic term in the string frame~(\ref{Reff}),  does not~\cite{Vollick:1998qf}, thereby playing a pivotal or effectively ``exotic''  role in supporting the wormhole geometry.}  In fact, the condition can be severely  violated    near  the singular points  $y=b_{\pm}$ and  in  the asymptotic  region  $(b_{+}+b_{-})y\rightarrow\infty$,  particularly when  $h^2<b^2$.

From the DFT perspective, however, the metric, $H$-flux, and dilaton are all gravitational components, forming the left-hand side of the DFT Einstein equation~(\ref{EDFE}). The matter part resides on the right-hand side,  \textit{i.e.~}$T_{AB}$,  which has its own energy conditions~\cite{Angus:2018mep,Angus:2019bqs} but is irrelevant to the  pure NS-NS geometry of the wormhole.    It is also worthwhile to note that, when converted to the Einstein frame, both the dilaton and the $H$-flux—viewed as ordinary matter—do not violate NEC.\vspace{3pt}

\section*{Traversable by not Particle but String} 
\begin{figure}[H]
\centering
\includegraphics[width=0.9\linewidth]{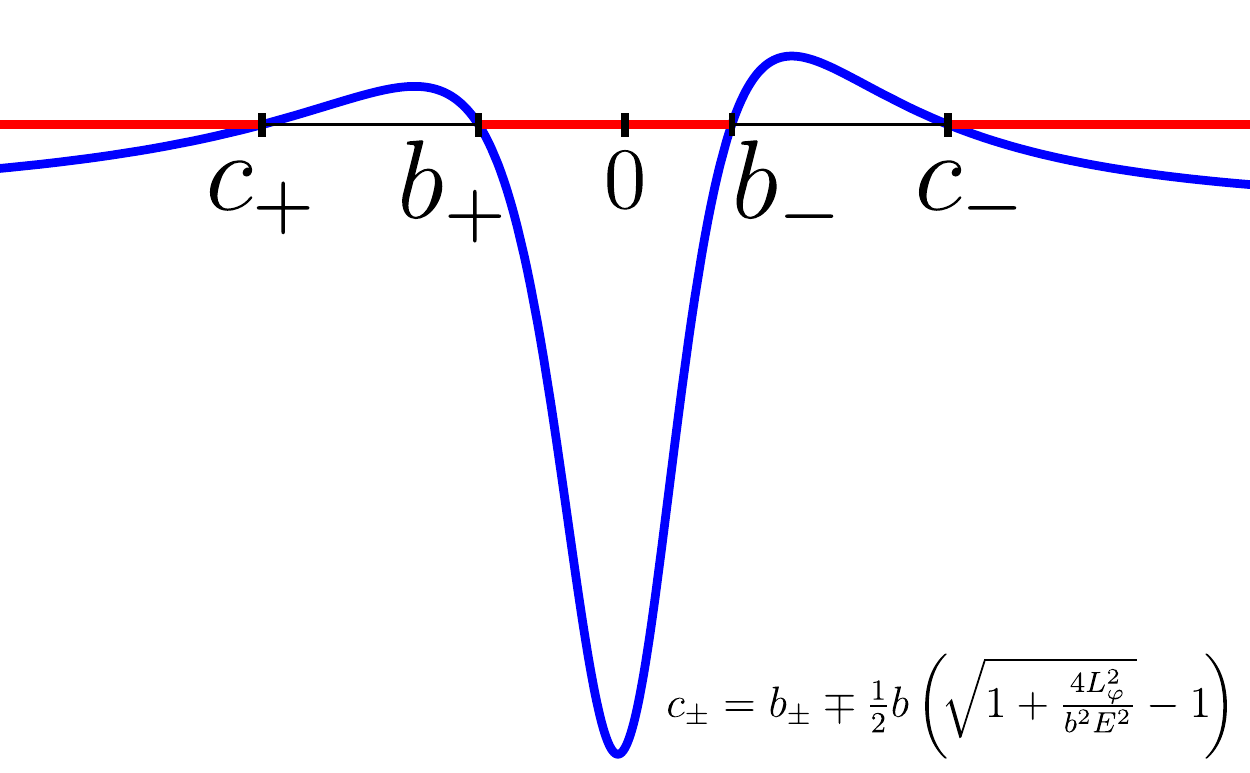}
\caption{{{Effective potential for null geodesics (${b>0}, {\Lp\neq0}$). Geodesics are complete and  confined in each of the three regions (red colored) divided by the points of  ${y=b_{\pm}}$.}}} 
	\label{FIGPotential}
\end{figure}

The semi-infinite region, ${y>\max(b_{+},b_{-})}$,  was demonstrated to be geodesically complete, without exhibiting any singular tidal force~\cite{Morand:2021xeq}. Here we extend the geodesic analysis to the full range, $y\in\mathbb{R}$.  We fix  ${\vartheta=\frac{\pi}{2}}$  without loss of generality and denote the conserved energy and angular momentum by constant  ${E\neq0}$ and $\Lp$. Null geodesic equations reduce to $\dot{t}=E\cF(y)$, $\dot{\varphi}=\Lp\cR(y)^{-2}$,  
and pivotally for the $y$-coordinate, 
\be
0=\dot{y}^{2}+V(y)\,,\quad\!\!\!\quad
V(y)=\left[-E^{2}\cF(y)+\frac{\Lp^{2}}{\cR(y)^{2}}\right]\cF(y)\,.
\label{0yV}
\ee
When $\Lp\neq0$, the effective potential $V(y)$ features two positive peak,  as depicted in FIG.\,\ref{FIGPotential}, such that a massless particle   cannot traverse ${y=b_{+}}$ nor ${y=b_{-}}$. If ${\Lp=0}$, it takes infinite amount of affine parameter, say $\lambda$, to reach the two points, as  $\scalebox{1.2}{$\int$}{\rd}\lambda=\scalebox{1.2}{$\int$}\!\frac{{\rd}y}{E\cF(y)}$ is logarithmically divergent.  In this way, the three regions divided by $y=b_{+}$ and $y=b_{-}$ are geodesically complete and the wormhole is non-traversable by  particles.\vspace{-7pt}
\vspace{-7pt}

We now turn to  strings.     
In terms of   light-cone coordinates   on  worldsheet,   $\sigma^{\pm}=\sigma\pm\tau$,  with conformal gauge,     the propagation of a string is  dictated  by
\be
\partial_{+}\partial_{-}x^{\mu}+\left(\Gamma^{\mu}_{\rho\sigma}+\frac{1}{2}H^{\mu}{}_{\rho\sigma}\right)\partial_{+}x^{\rho}\partial_{-}x^{\sigma}=0\,,
\label{sEoM1}
\ee
subject to   Virasoro constraints, $\partial_{\pm}x^{\mu}\partial_{\pm}x^{\nu}g_{\mu\nu}=0$.    An equivalent formula  to (\ref{sEoM1}) is 
\be
\scalebox{0.96}{$
\partial_{+}(g_{\mu\nu}\partial_{-}x^{\nu})\!+\!\partial_{-}(g_{\mu\nu}\partial_{+}x^{\nu})\!+\!(H_{\mu\nu\rho}{-\partial_{\mu}g_{\nu\rho}})\partial_{+}x^{\nu}\partial_{-}x^{\rho}=0\,.$}
\label{sEoM2}
\ee
{For the wormhole geometry~(\ref{SOL}),  we focus on  the radial propagation,  by  letting the two angular variables, $\vartheta,\varphi$ constant, and reduce  the  equation of motion~(\ref{sEoM2})  to
\be
\ba{r}
\partial_{+}\!\left[\frac{\partial_{-}t}{\cF(y)}\right]+\partial_{-}\!\left[\frac{\partial_{+}t}{\cF(y)}\right]=0\,,\\
\partial_{+}\!\left[\frac{\partial_{-}y}{\cF(y)}\right]+\partial_{-}\!\left[\frac{\partial_{+}y}{\cF(y)}\right]+\frac{\cF(y)^{\prime}}{\cF(y)^2}(\partial_{+}y\partial_{-}y-\partial_{+}t\partial_{-}t)
=0\,,
\ea
\label{EoM2}
\ee
and   the Virasoro constraints to
\be
\ba{ll}
(\partial_{+}y)^2-(\partial_{+}t)^2=0\,,\qquad&\qquad
(\partial_{-}y)^2-(\partial_{-}t)^2=0\,.
\ea
\label{Virasoro2}
\ee
To  obtain the  most general solutions to (\ref{EoM2}) and (\ref{Virasoro2}), we first define 
\be
\cG(y)=y\,+\textstyle{\left(\frac{b_{-}^{2}+{h^{2}/4}}{b}\right)}\ln\left|y-b_{-}\right|\,-\textstyle{\left(\frac{b_{+}^{2}+{h^{2}/4}}{b}\right)}\ln\left|y-b_{+}\right|\,,
\label{cGy}
\ee
which is the integral of $\cF(y)^{-1}$,  satisfying   
\be
\frac{\rd\cG(y)}{\rd y}=\frac{1}{\cF(y)}\,.
\label{dGyF}
\ee 
This provides useful identities:
\be
\partial_{+}\!\left[\frac{\partial_{-}y}{\cF(y)}\right]=\partial_{-}\!\left[\frac{\partial_{+}y}{\cF(y)}\right] =\partial_{+}\partial_{-}\cG(y)\,.
\label{uI}
\ee
Additionally, the simplified  Virasoro  constraints~(\ref{Virasoro2})  yield, with two independent sign factors $\pm$ and $\pm^{\prime}$, 
\be
\ba{ll}
\partial_{+}y=\pm\partial_{+}t\,,\qquad&\qquad \partial_{-}y=\pm^{\prime}\partial_{-}t\,,
\ea
\label{fourcases}
\ee
which  imply either   $\partial_{+}y\partial_{-}y=\partial_{+}t\partial_{-}t$ for  the same sign  or  $\partial_{+}y\partial_{-}y=-\partial_{+}t\partial_{-}t$ for opposite signs. These two cases correspond to \textit{i)} non-traversing and \textit{ii)} traversing solutions, respectively, as follows.

\textit{i)} When $\partial_{+}y\partial_{-}y=\partial_{+}t\partial_{-}t$,  we get ${y=\pm t}$, since   (\ref{fourcases}) with the same sign implies that  ${y\mp t}$ is constant (which we set trivial),  because $\partial_{+}(y\mp t)=0= \partial_{-}(y\mp t)$.  This solves (\ref{Virasoro2}) and, with the identities~(\ref{uI}),  reduces the two equations in (\ref{EoM2})  to
\be
\partial_{+}\partial_{-}\cG(y)=0\,,
\label{pmG}
\ee
whose  solution decomposes into   left- and right-movers,
\be
\cG(y)=\cG_{0}+2\alpha^{\prime}p\,\tau+f_{+}(\sigma^{+})+f_{-}(\sigma^{-})\,,
\label{cGyff}
\ee
where  $2\tau={\sigma^{+}-\sigma^{-}}$.  For a closed string $f_{+}(\sigma^{+})$ and $f_{-}(\sigma^{-})$ are arbitrary  periodic functions, leading to    vibrational mode expansions, whereas   an open string needs to meet   Neumann or Dirichlet boundary conditions~\cite{Green:1987sp}. In any case, equating (\ref{cGy}) and  (\ref{cGyff}), we arrive at non-chiral solutions, $y(\sigma^{+},\sigma^{-})$ and  $t(\sigma^{+},\sigma^{-})=\pm y(\sigma^{+},\sigma^{-})$,  in terms of the  inverse function  of $\cG(y)$, at least locally.  

  In particular, far away from the non-Riemannian  points  ${y=b_{\pm}}$, we have ${\cG(y)\simeq y}$~(\ref{cGy}) and thus---not surprisingly---the string propagates like a free string on a flat background.     However, such a  non-chiral string  cannot approach nor cross the points  ${y=b_{\pm}}$ with finite amount of $\tau$,  since from (\ref{cGy}), $\cG(y)$ would diverge but the right hand side of the equality in (\ref{cGyff}) ought to be finite.  Only in the limit, $\tau\rightarrow\pm\infty$, the string may reach  the non-Riemannian points. These observations are all consistent with the particle geodesics discussed above.  In fact,  from the perspective of the target spacetime, the string configuration $y=\pm t$ behaves indistinguishably from a point particle,  concealing its spatial extension.   Furthermore,  in a  point-particle limit of the non-chiral string---where the two functions 
   $f_{+}(\sigma^{+})$ and $f_{-}(\sigma^{-})$ become  constant---equations  (\ref{dGyF}) and (\ref{cGyff}) yield the relation, 
\be
2\alpha^{\prime}p=\frac{\frac{\rmd y}{\rmd\tau}}{\cF(y)}\,.
\ee 
This precisely reproduces the radial geodesic motion of a point particle described by (\ref{0yV}), with angular momentum ${L_{\varphi}=0}$ and energy $E=2\alpha^{\prime}p$.

\indent \textit{ii)} When  $\partial_{+}y\partial_{-}y=-\partial_{+}t\partial_{-}t$, (\ref{fourcases}) implies   
\be
\ba{ll}
\partial_{+}y=\pm \partial_{+} t\,,\qquad&\qquad \partial_{-}y=\mp \partial_{-}t\,.
\ea
\label{caseTWO}
\ee
Consequently, with the identities~(\ref{uI}),  the former equation in (\ref{EoM2})  holds trivially, while the latter one becomes
\be
\ba{rll}
0&=&\partial_{+}\partial_{-}\cG(y)+\frac{\cF(y)^{\prime}}{\cF(y)^2}\partial_{+}y\partial_{-}y\\
{}&=&\partial_{+}\left(\frac{1}{\cF}\partial_{-}y\right)-\left(\partial_{+}\frac{1}{\cF}\right)\partial_{-}y\,=\,\frac{1}{\cF}\partial_{+}\partial_{-}y\,,
\ea
\label{ppy}
\ee
which decomposes  $y$ into   left- and right-movers. The  full solutions to (\ref{caseTWO}) and (\ref{ppy}) are, \textit{c.f.~}(\ref{cGyff}),
\be
\ba{rll}
y&=&2\alpha^{\prime}p\,\tau+f_{+}(\sigma^{+})+f_{-}(\sigma^{-})\,,\\
\pm t&=&2\alpha^{\prime}p\,\sigma+f_{+}(\sigma^{+})-f_{-}(\sigma^{-})\,,
\ea
\label{transSOL}
\ee
such that ${y\pm t}$ is chiral and ${y\mp t}$ is anti-chiral,  like the non-relativistic string~\cite{Gomis:2000bd}.   

  When ${p\neq 0}$,  as  the worldsheet time $\tau$ evolves from $-\infty$ to $+\infty$, the chiral  string~(\ref{transSOL})  clearly traverses the wormhole along the $y$ direction.   Even when ${p=0}$, which is  required by the periodic boundary conditions   not only on spatial $y$ but also on temporal $t$,    the chiral string~(\ref{transSOL}) can still traverse the wormhole freely, provided the amplitudes of the periodic functions $f_{\pm}$ are arbitrarily  large.

In particular, if  we set either $f_{-}=\alpha^{\prime} p\,\sigma^{-}$ or  $f_{+}=-\alpha^{\prime} p\,\sigma^{+}$  in (\ref{transSOL}),  we obtain  chiral or anti-chiral solutions:
\be
\ba{ll}
y=\pm t=\alpha^{\prime} p\sigma^{+}{+f_{+}(\sigma^{+})}\,,\quad&~ y=\mp t=-\alpha^{\prime} p\sigma^{-}{+f_{-}(\sigma^{-})}\,.
\ea
\label{ytchiral}
\ee
In this scenario, the chiral string behaves as if it were “point-like,” in the sense that it exhibits  no spatial extension, \textit{c.f.~}\cite{Park:2020ixf,Jusinskas:2021bdj,Lize:2021una}.   Nonetheless,  unlike   a genuine point particle~(\ref{0yV}), this chiral string can traverse the wormhole.     

Alternatively,  if we choose  an identical  function $f_{+}(\sigma)=f_{-}(\sigma)=b\sin\sigma$ with ${p=0}$ in (\ref{transSOL}),  we get  $y=2b\cos\tau\sin\sigma$  and $t=2b\sin\tau\cos\sigma$.  The string forms  an ellipsoid on the target spacetime which  encompasses the wormhole,\vspace{-5pt}
\be
\Big(\frac{t}{\cos\sigma}\Big)^{2}+\Big(\frac{y}{\sin\sigma}\Big)^{2}=4b^2\,.
\label{ellipsoid}
\ee
This configuration is spatially extended across the wormhole geometry, with the spatial coordinate $y$  parametrized by $\sigma$ at each fixed time $t$.  }

\section*{Conclusion} 
{We have identified a  wormhole traversable by string, purely within the NS-NS sector, without invoking any exotic matter.} {If string theory is realized in Nature—where all matter is composed of tiny vibrating strings—our findings demonstrate that analyzing wormhole traversability solely in terms of point-particle geodesics is inadequate.}

Notably, the traversing chiral-string solution~(\ref{transSOL}) appears to transcend the specific details of the wormhole geometry, \textit{e.g.~}$\cF(y)$, hinting at the pregeometric nature of chiral strings. Nonetheless, removing the $H$-flux from (\ref{sEoM1}) would invalidate the traversability {which 
 itself  supports the interpretation of the points ${y=b_{\pm}}$ as DFT regularity rather than GR singularity.}

We conjecture that an ordinary string near a wormhole may split into chiral and anti-chiral modes to traverse it. After crossing, these modes could merge back into a single ordinary string.\\

\noindent{\textit{Acknowledgments.}}---We  wish to thank Wonwoo Lee   for  useful comments.  This  work is supported by the National Research Foundation of Korea (NRF)  through Grants: NRF-2023R1A2C2005360, NRF-2023K2A9A1A01098740,  NRF-2022R1I1A1A01069032,  and  RS-2020-NR049598  (Center for Quantum Spacetime: CQUeST).\\

\noindent{\textit{E-mail}}: \textsf{hun.jang@nyu.edu,  mkim@sogang.ac.kr, insaying@sogang.ac.kr,  park@sogang.ac.kr}\\
\hfill


\appendix
\renewcommand\thesection{A}
\renewcommand{\thefigure}{A\,\arabic{figure}}
\setlength{\jot}{9pt}                 
\renewcommand{\arraystretch}{3.2} 
\setcounter{equation}{0}
\setcounter{figure}{0}

\begin{center}
\vspace{9pt}
	\Large	\textbf{APPENDIX}
\end{center}
\vspace{5pt}


\section*{String's Equations of Motion}
In general, taking the partial derivatives, $\partial_{+}$ or $\partial_{-}$, on the Virasoro constraints,
\be
\ba{ll}
 \partial_{+}x^{\mu}\partial_{+}x^{\nu}g_{\mu\nu}=0\,,\qquad&\qquad
 \partial_{-}x^{\mu}\partial_{-}x^{\nu}g_{\mu\nu}=0\,,
 \ea
 \ee
 one obtains
 \be
 \ba{r}
\partial_{+}x^{\lambda}g_{\lambda\mu}\left(\partial_{+}^2x^{\mu}+\Gamma^{\mu}_{\rho\sigma}\partial_{+}x^{\rho}\partial_{+}x^{\sigma}\right) =0\,,\\
\partial_{-}x^{\lambda}g_{\lambda\mu}\left(\partial_{-}^2x^{\mu}+\Gamma^{\mu}_{\rho\sigma}\partial_{-}x^{\rho}\partial_{-}x^{\sigma}\right) =0\,,
\ea
\label{furtherpartial1}
\ee
and
\be
\ba{r}
\partial_{+}x^{\lambda}g_{\lambda\mu}\left(\partial_{+}\partial_{-}x^{\mu}+\Gamma^{\mu}_{\rho\sigma}\partial_{+}x^{\rho}\partial_{-}x^{\sigma}\right) =0\,,\\
\partial_{-}x^{\lambda}g_{\lambda\mu}\left(\partial_{+}\partial_{-}x^{\mu}+\Gamma^{\mu}_{\rho\sigma}\partial_{+}x^{\rho}\partial_{-}x^{\sigma}\right) =0\,.
 \ea
 \label{furtherpartial2}
 \ee
 The  string's equation of motion~(\ref{sEoM1}) which can be reformulated into
 \be
 g_{\lambda\mu}\left(\partial_{+}\partial_{-}x^{\mu}+\Gamma^{\mu}_{\rho\sigma}\partial_{+}x^{\rho}\partial_{-}x^{\sigma}\right)=-{\frac{1}{2}}H_{\lambda\rho\sigma}\partial_{+}x^{\rho}\partial_{-}x^{\sigma}\,,
 \ee
 is then consistent with  (\ref{furtherpartial2}), due to the skew-symmetric property of the $H$-flux, while (\ref{furtherpartial1}) remains   independent.  
 
\begin{widetext} 
For the wormhole geometry~(\ref{SOL}),  the full set of string's equations of motion   are  explicitly:

\be
\ba{r}
\partial_{+}\!\left[\frac{\partial_{-}t}{\cF(y)}\right]+\partial_{-}\!\left[\frac{\partial_{+}t}{\cF(y)}\right]+h\sin\vartheta (\partial_{-}\vartheta\partial_{+}\varphi-\partial_{+}\vartheta\partial_{-}\varphi)=0\,,\\
\partial_{+}\!\left[\frac{\partial_{-}y}{\cF(y)}\right]+\partial_{-}\!\left[\frac{\partial_{+}y}{\cF(y)}\right]+\frac{\cF(y)^{\prime}}{\cF(y)^2}(\partial_{+}y\partial_{-}y-\partial_{+}t\partial_{-}t)
-2y(\partial_{+}\vartheta\partial_{-}\vartheta+\sin^2\vartheta\partial_{+}\varphi\partial_{-}\varphi)=0\,,\\
\partial_{+}\!\left[\cR(y)^2\partial_{-}\vartheta\right]+\partial_{-}\!\left[\cR(y)^2\partial_{+}\vartheta\right]-2\cR(y)^2\sin\vartheta\cos\vartheta\partial_{+}\varphi\partial_{-}\varphi
+h\sin\vartheta (\partial_{+}\varphi\partial_{-}t-\partial_{-}\varphi\partial_{+}t)=0\,,\\
\partial_{+}\left[\cR(y)^2\sin^2\vartheta\partial_{-}\varphi\right]+\partial_{-}\left[\cR(y)^2\sin^2\vartheta\partial_{+}\varphi\right]+h\sin\vartheta (\partial_{+}t\partial_{-}\vartheta-\partial_{-}t\partial_{+}\vartheta)=0\,,
\ea
\label{EoMSM2}
\ee
and  the Virasoro constraints are:
\be
\ba{ll}
\scalebox{1.2}{$\frac{(\partial_{+}y)^2-(\partial_{+}t)^2}{(y-b_{+})(y-b_{-})}$}+(\partial_{+}\vartheta)^2+\sin^2\vartheta\,(\partial_{+}\varphi)^2=0\,,\qquad&\qquad
\scalebox{1.2}{$\frac{(\partial_{-}y)^2-(\partial_{-}t)^2}{(y-b_{+})(y-b_{-})}$}+(\partial_{-}\vartheta)^2+\sin^2\vartheta\,(\partial_{-}\varphi)^2=0\,.
\ea
\label{VirasoroSM2}
\ee
\end{widetext}
\section*{$\ODD$-symmetric Volume of the Wormhole}
The integral  of the $\ODD$-symmetric volume form, $e^{-2d}$~(\ref{PARA}),   over  the middle throat region, say $\Sigma_{t}$ (a slice of constant $t$), coincides with  the three sphere volume of radius $b$, hence independent of the $H$-flux:
\be
\int_{\Sigma_{t}}e^{-2d}=
\int_{b_{+}}^{b_{-}}\!\rd y\int_{0}^{\pi}\!\!\rd\vartheta\int_{0}^{2\pi}\!\!\rd\varphi~\cR(y)^2\sin\vartheta=\frac{4\pi}{3}b^3\,.
\ee
Given that the throat has the minimal area ${4\pi\cR(0)^2=\pi h^2}$, the (averaged) height of the middle    throat region is, roughly speaking, inversely proportional to the $H$-flux squared,  $\propto{b^3}/{3h^2}$. It remains to be seen what would be the (flat spacetime) holographic interpretation~\cite{VanRaamsdonk:2010pw}, regarding   \textit{e.g.~}complexity~\cite{Susskind:2014moa},  if any. \\



~\\
~\\

\end{document}